\documentclass[a4paper,7pt]{article}
\usepackage{authblk}
\usepackage[dvips]{graphicx}
\graphicspath{{noiseimages/}}
\usepackage{xcolor}
\usepackage{tikz}
\usepackage{pgfplots}
\usetikzlibrary{intersections, pgfplots.fillbetween}
\usetikzlibrary{decorations.pathmorphing}
\usepackage{amsmath}
\usepackage{amssymb}
\usepackage{a4wide}
\usepackage[english]{babel}
\usepackage{amsmath}
\usepackage{hyperref}
\usepackage[toc,page]{appendix}
\definecolor{urlcolor}{HTML}{990000}
\definecolor{linkcolor}{HTML}{005F5F}
\hypersetup{pdfstartview=FitH,  linkcolor=linkcolor,urlcolor=urlcolor, colorlinks=true,citecolor=blue}
\newcommand{\tanhx}{u}
\newcommand{\ind}{{-\frac12+i\mu}}

\renewcommand{\mu}{{\nu}}

\renewcommand{\Im}{\operatorname{Im}}
\def\bear{\begin{eqnarray}}
\def\ear{\end{eqnarray}\noindent}
\newcommand{\sech}{{\rm{sech\,}}}
\def\2F1{\phantom{}_2\hspace{-.95pt}F_1}
\author[1,2]{E.T.Akhmedov}
\author[1,2]{K.V.Bazarov}
\author[1,2]{D.V.Diakonov}
\author[3,4]{U.Moschella}
\affil[1]{Institutskii per. 9, Moscow Institute of Physics and Technology, 141700, Dolgoprudny, Russia}
\affil[2]{B. Cheremushkinskaya, 25, Institute for Theoretical and Experimental Physics, 117218, Moscow, Russia}
\affil[3]{Universit\`a degli Studi dell'Insubria - Dipartimento DiSAT, Via Valleggio 11 - 22100 Como - Italy}
\affil[4]{INFN, Sez di Milano, Via Celoria 16, 20146, Milano - Italy}
\title{\textcolor{black}{Quantum fields in the static de Sitter universe}}

\begin{document}

\numberwithin{equation}{section}

\maketitle

\begin{abstract}

We construct explicit mode expansions of various tree-level propagators in the Rindler -- de Sitter universe, also known as  the static (or compact) patch of the de Sitter spacetime. We construct in particular the Wightman functions for  thermal states having a  generic temperature $T$. We give a fresh simple proof that the only thermal Wightman propagator
that respects the de Sitter isometry is the restriction to the Rindler -- de Sitter wedge of the propagator for the Bunch--Davies state. It is the thermal state with $T = (2 \pi)^{-1}$ in the units of de Sitter curvature. We show that propagators with $T\ne(2\pi)^{-1}$ are only time translation invariant and  have extra singularities on the boundary of the static patch.
We also construct the expansions for the so-called alpha-vacua in the static patch and discuss the flat limit.

\end{abstract}
\newpage

\tableofcontents

\newpage

\section{Introduction}

Notwithstanding  the existence of a vast literature on   de Sitter quantum fields there is still no consensus as regards their infrared behaviour which is very much different from the behaviour of  Minkowski and anti de Sitter fields.
The root of the problem is the absence of a  globally defined timelike Killing vector field on the de Sitter manifold: the components of the metric in general depend on the  time variable $t$ of the chosen coordinate system.

The non-stationarity of the de Sitter metric indicates that to compute loop quantum  corrections  one should make use of the Schwinger-Keldysh  rather than the Feynman diagrammatic technique;
in the first step, initial data have to be imposed on a Cauchy surface at some  time $t_0$ to define the correlation functions. No matter what state is chosen, secular growing infrared contributions  appear in the loops  (see e.g. \cite{Akhmedov:2013vka} for a review);
these   effects  are global and sensitive to the initial  conditions \cite{Akhmedov:2013vka,Akhmedov:2019cfd,Akhmedov:2017ooy,Akhmedov:2019esv}.
Summarizing, when quantizing fields in curved space--times, the field dynamics may and in general does  depend on the choice of coordinates through the choice of the initial data and this is crucial, in particular, for understanding the properties of de Sitter quantum physics.

Cosmologists usually make use  either of the spatially flat Poincar\'e  coordinates  \cite{lem,sch,mosch}
or of the global spherical coordinates  \cite{sch,mosch,Lanczos} for the de Sitter manifold. To build  correlation functions
the common initial choice is the Bunch--Davies (also called Euclidean) de Sitter invariant state \cite{thir,  nach,Chernikov, ss, Bunch,GH}. What makes it special is that this is the only state  among the de Sitter invariant ones -- the so  called alpha vacua \cite{spindel, Mottola, Allen} -- to be maximally analytic \cite{bm,bgm}.

There is however a particular chart -- the  "static patch"   discovered by de Sitter in 1917 \cite{desitter2} -- that admits a timelike Killing vector field. The field of course is not globally timelike: it becomes lightlike on the boundary of the static patch (the horizons) and spacelike beyond it.
A celebrated result by Gibbons and Hawking\footnote{This result was actually predated  by another important result by Figari, Hoegh-Krohn and Nappi \cite{fhkn} who studied interacting quantum fields in the wedge in two dimensions by applying constructive methods on the Euclidean sphere. }
\cite{GH,bm,bgm,sewell,nthir} says that {\em the restriction to the static patch of the Bunch-Davies state is a thermal equilibrium state at the temperature $1/2\pi R$ w.r.t. the relevant time coordinate}, where $R$ is the de Sitter radius.

As one can see, it is a very specific statement that  (of course) is true only when all the above conditions are fulfilled. Nonetheless it is not uncommon to hear or read the (incorrect) general  statement that "the de Sitter space has a temperature" which is sometimes taken as the starting point of vague speculations about cosmology and quantum gravity.

Apart from the Gibbons--Hawking result, not very much is known about quantum fields in the Rindler -- de Sitter wedge (static patch).  The relevance of this model for cosmology and black hole physics makes this lack of knowledge is even more surprising.

In this paper we partially fill this gap by constructing all the time translation invariant states at tree--level. In particular we produce new (to the best of our knowledge) formulae giving  explicit mode expansions of the correlation functions where the coordinates of the static patch are separated. This is an obviously necessary preliminary step to apply perturbation theory and calculate loops as described above.

Our class of equilibrium states  contains in particular all the thermal states and we give a new direct proof that the complete de Sitter invariance is recovered only at $T= 1/2\pi R$.
All the other states, including the zero temperature vacuum (pure) state, are not de Sitter invariant and have unusual singularities at the horizon, giving retrospectively some support to Einstein's suspicions about the equator of the patch \cite{janssen}.

%In particular it seems to mean that, unlike the Minkowski space case, one cannot heat up or cool down a field theory in de Sitter space to an arbitrary temperature. Or prepare a state with an arbitrary temperature at a given moment of time. These observations one can make already at the tree--level.

We restrict our attention to the two--dimensional case to avoid unnecessary complications in the equations. Most of our results can be straightforwardly extended to other dimensions. Loop corrections to tree--level propagators will be investigated in a companion paper.

\section{Geometry}
\label{geometry}
The coordinate system which is of interest for us here was introduced as early as 1917 by Willem de Sitter  in the course of the famous debate on the relativity of inertia \cite {janssen}.

\begin{figure}[h]
    \centering
    \includegraphics[scale=0.8]{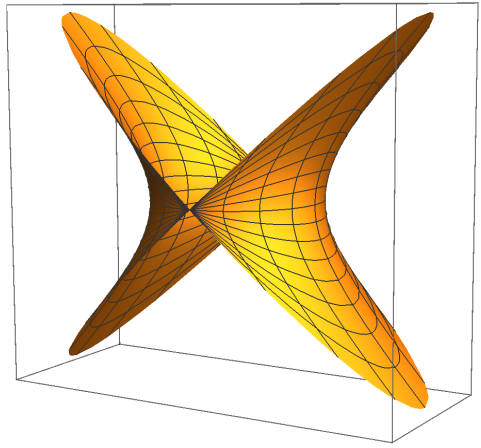}
    \caption{The static patch and it opposite seen as right and left Rindler-de Sitter wedges.}
    \label{fig:spatch}
\end{figure}

%A century after the Einstein -- de Sitter debate the role of coordinates in General Relativity and Cosmology is much better understood. On the other hand,  when quantizing fields in curved space--times (and, even worse, fields and gravity together) things are not yet so clear, because the field dynamics may depend on the choice of  coordinates, as we have mentioned above and will comment more below in this section.
By visualizing the two-dimensional de Sitter space as the one-sheeted  hyperboloid embedded in a three dimensional ambient Minkowski space
\begin{align} \label{dssa}
dS_2 = \{X\in {\bf R}^3, \ \  X^\alpha X_\alpha=X_{0}^2-X_{1}^2-X_{2}^2=-R^2\} \qquad %ds_3^2=dX_\mu dX^{\mu}
\end{align}
(capital $X^{\alpha}$ denote the coordinates of a given Lorentzian frame  of the ambient spacetime) the de Sitter static coordinates are
\begin{equation}
\label{coordinates}
 X\left(\frac t R,\frac x R\right)=\begin{cases}
   X^0= R \sinh \frac t R \ \sech \frac x  R\\
   X^1=R \tanh \frac x R = u \\
   X^2= R \cosh \frac t R \ \sech\frac  x R
 \end{cases},
 \qquad t \in(-\infty,\infty), \ x \in(-\infty,\infty).
\end{equation}
In the following we will set $R=1$ and $\tanh x = \tanhx$.

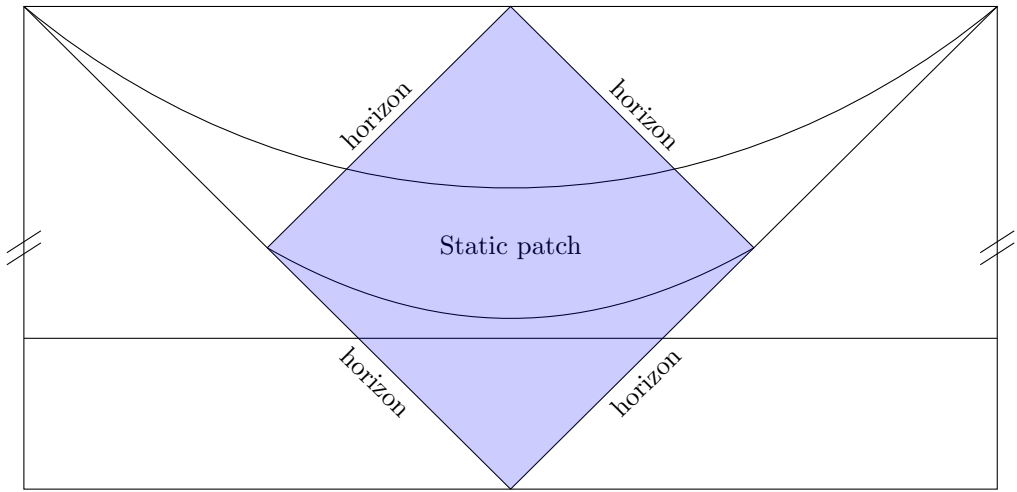
\begin{figure}[h!]
\centering
\begin{tikzpicture}[scale=0.8]
\node (I)    at ( 4,0)   {};
\node (II)   at (-4,0)   {Static patch};
\node (III)   at (-12,0)   {};
\node (IV)   at (0,-2.5) {};

%additional lines
\node (Lt) at (-12,1/10)  {};
\node (Lb) at (-12,-1/10)  {};

%globalsurface
\coordinate (gl) at (-12,-3/2)  {};
\coordinate (gr) at (4,-3/2)  {};
\draw (gl) -- (gr);

%patch
\coordinate (pl) at (-12,4)  {};
\coordinate (pr) at (4,4)  {};
\draw (pl) to[out=-40,in=-140] (pr);
\coordinate (sr) at (0,0)  {};
\coordinate (sl) at (-8,0)  {};
\draw (sl) to[out=-30,in=-150] (sr);

\path  % Four corners of left additional line
  (Lt)
       +(0+33:1/3)   coordinate    (Ltright)
       +(180+33:1/3) coordinate (Ltleft)
       ;

\path  % Four corners of left additional line
  (Lb)
       +(0+33:1/3)   coordinate    (Lbright)
       +(180+33:1/3) coordinate (Lbleft)
       ;

\draw (Ltright) -- (Ltleft);
\draw (Lbright) -- (Lbleft);

\node (Rt) at (4,1/10)  {};
\node (Rb) at (4,-1/10)  {};

\path  % Four corners of left additional line
  (Rt)
       +(0+33:1/3)   coordinate    (Rtright)
       +(180+33:1/3) coordinate (Rtleft)
       ;

\path  % Four corners of left additional line
  (Rb)
       +(0+33:1/3)   coordinate    (Rbright)
       +(180+33:1/3) coordinate (Rbleft)
       ;

\draw (Rtright) -- (Rtleft);
\draw (Rbright) -- (Rbleft);

%diamond

\path  % Four corners of left diamond
  (II) +(90:4)  coordinate  (IItop)
      % +(-90:4) coordinate[label=-90:$t\to -\infty$] (IIbot)
       +(-90:4) coordinate (IIbot)
       +(0:4)   coordinate    (IIright)
       +(180:4) coordinate (IIleft)
       ;
\draw[name path = A](IIleft) --
          node[midway, above left]    {}
          node[midway, above, sloped] {horizon}
      (IItop) --
          node[midway, above, sloped] {horizon}
      (IIright) ;

\draw[name path = B](IIright) --
          node[midway, below, sloped] {horizon}
      (IIbot) --
          node[midway, below, sloped] {horizon}
          node[midway, below, left]    {}
      (IIleft);

\tikzfillbetween[of=A and B]{blue, opacity=0.2};

\path % Four conners of the right diamond (no labels this time)
   (I) +(90:4)  coordinate (Itop)
       +(-90:4) coordinate (Ibot)
       +(180:4) coordinate (Ileft)
       ;

% No text this time in the next diagram
%\draw  (Ileft) -- (Itop) -- (Ibot) -- (Ileft) -- cycle;

\path % Four conners of the left diamond (no labels this time)
   (III) +(90:4)  coordinate (IIItop)
        +(-90:4) coordinate (IIIbot)
       ;

% No text this time in the next diagram
\draw (IIbot) -- (IIIbot) -- (IIItop) -- (IItop);
\draw (IIItop) -- (IIleft);
\draw (Itop) -- (IIright);
\draw (IIbot) -- (Ibot) -- (Itop) -- (IItop);
%% Squiggly lines
%\draw[decorate,decoration=zigzag] (IItop) -- (Itop)
%      node[midway, above, inner sep=2mm] {$r=0$};
%
%\draw[decorate,decoration=zigzag] (IIbot) -- (Ibot)
%     node[midway, below, inner sep=2mm] {$r=0$};
\end{tikzpicture}
\caption{Penrose diagram of the de Sitter manifold with Cauchy surfaces of different patches. The static patch is bordered by a bifurcate Killing horizon.}
\label{dssp}
\end{figure}

From a group theoretical viewpoint
the new time coordinate $t$ parametrizes the one-parameter subgroup of the de Sitter group stabilizing the equator. The action of that subgroup to points of any spherical spatial section containing the equator gives coordinates to two opposite static patches.  In two dimensions the spatial sections are  ellipses; the equator degenerates in the two points where all the ellipses meet (see Fig. \ref{fig:spatch}).
A static patch is in fact the intersection of the de Sitter manifold with a Rindler wedge in one dimension more and  this is why we also call it  as Rindler -- de Sitter wedge. The above coordinates thus cover only  the region $\{|X^1|<1\}  \cap
\{X^2>|X^0|\}$ of the {\em real} de Sitter manifold, the right wedge in Fig. \ref{fig:spatch},  the shaded region in Fig. \ref{dssp}.

A Rindler -- de Sitter wedge is itself a globally hyperbolic space--time but a Cauchy surface for the wedge is incomplete w.r.t. the whole de Sitter manifold, being only "one half" of a bona fide Cauchy surface\footnote{In saying this we are {\em supposing} that the geodesical completion of the wedge is the de Sitter manifold. Would we {\em suppose} that the geodesical completion be, say, its double covering, the result would change completely. In particular there would be no thermal state at all \cite{Epstein2020}.} , see Figs. \ref{fig:spatch} and \ref{dssp}.  On the other hand quantization in the static coordinates has an advantage in comparison with other coordinate systems:  the Hamiltonian operator is time independent.
The metric
\begin{align}
\label{metric}
ds^2 = \frac{dt^2-dx^2}{\cosh x {}^2}
\end{align}
 is time independent and conformal to the flat metric.
The  static patch is bordered by a bifurcate Killing horizon
\begin{align*}x\to \pm \infty, \ \ \  t=\pm x %+const.
\end{align*}
where the metric degenerates.
The corresponding Killing vector is not time-like when extended outside the static patch. The de Sitter  invariant scalar product is given by
\begin{gather}
\label{Zdist}
    \zeta = \zeta_{12} = X_1^{\alpha}{X_{2}}_\alpha=- \frac{\cosh(t_1 - t_2) + \sinh x_1 \sinh x_2}{\cosh x_1 \cosh x_2}.
\end{gather}
The geodesic distance $L$ and $\zeta$ are related as follows:
$\zeta = - \cosh(L)$  \text{for time-like geodesics},   $\zeta = \cos(L)$ \text{for space-like ones};  $\zeta = - 1$ for light-like separations or coincident points.
%\section{Constructing Wightman function}
\section{Canonical Quantization}
\label{quantize}

In this section we outline the canonical
quantization of the Klein-Gordon field in the static chart coordinates:
\begin{eqnarray}
\label{sdseq}
&& \left(\partial_{t}^2 - \partial_x^2 + \frac{m^2}{\cosh^2 x}\right) \phi(t,x)=0,
\\
&& \label{cancom}\Big[ {\phi}(t,x_1),{{\phi}}(t,x_2) \Big] = 0, \ \ \ \
\Big[ {\phi}(t,x_1),\dot{{\phi}}(t,x_2) \Big] = i\delta(x_1-x_2).
\end{eqnarray}
The static chart is in itself a globally hyperbolic manifold, though geodesically incomplete. We may apply standard methods of canonical quantization and look for a complete set of modes by separating the variables. Of course the so constructed set of modes will be incomplete when considered w.r.t. the whole de Sitter manifold \cite{schaeffer,sch2}.

Let us  consider factorized modes which have  {\em positive frequencies } w.r.t. the time coordinate $t$:
\begin{equation}
\label{generalsol}
\varphi(t,x) = e^{-i\omega t} \psi_\omega(u), % = e^{-i\omega t} \Big\{ A_\omega  \mathsf{P}^{i\omega}_{-\frac12+ i\mu}(\tanhx)+B_\omega  \mathsf{P}^{i\omega}_{-\frac12+ i\mu}(-\tanhx) \Big\},
\quad \tanhx = \tanh x.
\end{equation}
$\psi_\omega(u)$ are  eigenfunctions of the continuous spectrum of the well-known quantum mechanical scattering problem:
\begin{align}
\bigg[- \partial_x^2 + \frac{ m^2}{\cosh^2 x}\bigg]\, \psi_{\omega}(u)=\omega^2 \psi_{\omega}(u), \quad \tanhx = \tanh x, \ \  \ \ m^2 = \frac{1}{4}+\nu^2. \label{QM}
\end{align}
For any given $\omega\geq 0$  the  Ferrers functions $\mathsf{P}^{i\omega}_{-\frac{1}{2}+i\mu}(\pm\tanhx)$ -- also known as  Legendre functions on the cut \cite{Bateman} --  are two independent solutions of the above equation.
 The  double degeneracy of the energy level $\omega$
points towards the introduction of two pairs of creation and annihilation operators for each level:
\begin{align}
\big[a_{\omega_1}, a^\dagger_{\omega_2} \big]=\delta(\omega_1-\omega_2), \qquad \big[b_{\omega_1}, b^\dagger_{\omega_2} \big]= \delta(\omega_1-\omega_2), \qquad \big[a_{\omega_1}, b_{\omega_2} \big]=\big[a_{\omega_1}, b^\dagger_{\omega_2} \big]=0.
\end{align}
The mode expansion of the field operator ${\phi}(t,x)$ can then be written as follows:
\begin{equation}
\label{fieldoperator}
{\phi}(t,x)=\int_{0}^\infty  \frac{d\omega}{2\pi} \bigg[ e^{-i \omega t} \Big(\psi_\omega(u) a_\omega+\psi_\omega(-u) b_\omega \Big)+e^{i \omega t} \Big(\psi^*_\omega(u) a^\dagger_\omega+\psi^*_\omega(-u) b^\dagger_\omega \Big)\bigg]
\end{equation}
where %$\psi_\omega(x)$ %is the spatial part of solution of \eqref{generalsol} with the proper normalization coefficient:
\begin{align}
\label{xmodes}
\psi_\omega(u)=\sqrt{\sinh(\pi\omega)} \  \Gamma\Big(\frac{1}{2}+i\mu-i\omega\Big)\, \Gamma\Big(\frac{1}{2}-i\mu-i\omega \Big)\mathsf{P}^{i\omega}_{-\frac{1}{2}+i\mu}(\tanhx).
\end{align}
The normalization has been chosen according with the completeness relation (\ref{compl2}) shown in  Appendix A.
At large positive $x$   the wave \begin{equation}
\psi_\omega(\tanh x)\sim e^{i\omega x} \ \ \ x\to\infty \end{equation} is purely right moving (at large negative $x\to -\infty$  the wave $\psi_\omega(-\tanh x)\sim e^{-i\omega x} $ is purely left moving).

%\textcolor{red}{Note that any two (correctly normalized) linear combinations of the above modes will provide  equivalent quantizations as long as the time dependence of the modes remains unchanged.}

By normal ordering w.r.t. the vacuum of the $a_{\omega}$ and $b_{\omega}$ operators  we get the free Hamiltonian in the standard form
 \begin{align}
 \label{hamiltionian}
:H: =\int_{-\infty}^{+\infty} dx \sqrt{g} \, : T^0_{\,0} : = \int_0^{+\infty} d\omega \ \omega \   \Big( {a}_{\omega}^{\dag}{a}_{\omega}+ {b}^{\dag}_{\omega}{b}_{\omega} \Big).
  \end{align}
Note that the range of integration over $\omega$ starts from zero (rather than $m$ as for a massive field in flat space). This is because   the "mass" term in the action %$(\ref{action})$
  \begin{align*}
  S_m=\int  d^2x \sqrt{g} \  m^2   \varphi^2(t,x),
\end{align*}
vanishes near the horizon  (recall that $\sqrt{g}=(\cosh x)^{-2}$).

\subsection{Thermal two-point functions}
The quantum mechanical  average over a thermal state of inverse temperature $\beta$ is given by
\begin{align}
\label{tav}
\left< {\cal O} \right>_\beta =\frac{\text{Tr}  \ {\rho} \, {\cal O} }{\text{Tr}  \ {\rho}} , \qquad {\rho} \equiv  e^{-\beta {H}}.
\end{align}
Although the previous expression is ill-defined in quantum field theory, it still allows to compute  the thermal two-point function at inverse temperature $\beta$ by assuming the Bose-Einstein distribution of the energy levels
\begin{equation}
    \left< {a}_{\omega}^{\dag} {a}_{\omega'}^{}  \right>_\beta =\left<  {b}_{\omega}^{\dag} {b}_{\omega'}^{} \right>_\beta = ({e^{\beta \omega }-1})^{-1}\delta(\omega-\omega').
\end{equation}
Eqs. \eqref{fieldoperator} and  \eqref{hamiltionian} give the following expression:
% \begin{multline}
% \label{wightman0}
%  W_\beta(t_1-t_2,\, x_1,x_2)= \Big< {\phi}(t_1,x_1){\phi}(t_2,x_2) \Big> =\\= \int_{0}^{\infty}\frac{d\omega  d\omega'}{4\pi^2}   \Bigg[  e^{-i\omega t_1+i\omega' t_2} \psi_{\omega}(x_1) \psi_{\omega'}^*(x_2)\Big< {a}_{\omega}^{} {a}_{\omega'}^{\dag} \Big> +
 % e^{i\omega t_1-i\omega' t_2} \psi^*_{\omega}(x_1) \psi_{\omega'}(x_2)\Big<{a}^{\dag}_{\omega} {a}_{\omega'}^{}\Big>+
 % \\ +e^{-i\omega t_1+i\omega' t_2} \psi_{\omega}(-x_1) \psi_{\omega'}^*(-x_2)\Big< {b}_{\omega}^{} {b}_{\omega'}^{\dag} \Big> +
 % e^{i\omega t_1-i\omega' t_2} \psi^*_{\omega}(-x_1)\psi_{\omega'}(-x_2)\Big<\hat{b}^{\dag}_{\omega} \hat{b}_{\omega'}^{}\Big>\Bigg].
 %\end{multline}
 %
  \begin{eqnarray}
 %\label{wightman0}
   W_\beta(t_1-t_2,\, x_1,x_2)&=& \left< {\phi}(t_1,x_1){\phi}(t_2,x_2) \right>_\beta = \int_{0}^{\infty}\frac{d\omega  }{4\pi^2}   \left[   \frac{e^{-i\omega (t_1-t_2)}}{1-e^{-\beta \omega }} \bigg(\psi_{\omega}(u_1) \psi_{\omega}^*(u_2)\right. \cr &+& \left. \, \psi_{\omega}(-u_1) \psi_{\omega}^*(-u_2)\bigg)  +
   \frac{e^{i\omega( t_1- t_2)}}{e^{\beta \omega }-1} \bigg(\psi^*_{\omega}(u_1) \psi_{\omega}(u_2)+\psi^*_{\omega}(-u_1)\psi_{\omega}(-u_2)\bigg)\right]  \cr  \cr
 \label{wightman}
 &=& \int_{-\infty}^{\infty} {e^{-i\omega (t_1-t_2)} }\frac{{1-e^{-2\pi\omega}}}{{1-e^{-\beta \omega}}}\widetilde P_{\nu}(\omega, u_1,u_2)\ d\omega \end{eqnarray}
 where
 \begin{equation}
\widetilde P_{\nu}(\omega,u_1,u_2) =
\frac { e^{\pi\omega} \left( \mathsf{P}^{i \omega}_{-\frac12 +i \nu}(u_1) \mathsf{P}^{-i \omega}_{-\frac12 -i \nu}(u_2)    + \mathsf{P}^{i \omega}_{-\frac12 +i \nu}(-u_1)
\mathsf{P}^{-i \omega}_{-\frac12 -i \nu}(-u_2) \right) }{ {8} \cosh\pi(\nu-\omega)\, {\cosh\pi(\nu+\omega)}}.
\end{equation}
The states defined by the above two-point functions are mixed. The only pure state is obtained in the limit $\beta \to \infty$.

In section \ref{bd8}  we will prove that for $\beta=2\pi$ the above two-point function is de Sitter invariant and coincides with the restriction to the static patch of the Bunch-Davies two-point function:
\begin{eqnarray}
  W_{2\pi}(t_1-t_2,\, x_1,x_2)= W_{BD} (\zeta) =
  \frac{1}{4\cosh \pi\nu} \, P_\ind(\zeta)\label{2DBDpropagator},
\end{eqnarray}
where $\zeta$ is the de Sitter invariant variable  defined in (\ref{Zdist}).
On the other hand for arbitrary $\beta$ the two--point function $ W_{\beta}(t_1-t_2, x_1,x_2)$ and its permuted function do not respect the de Sitter isometry because their  periodicity thermal property in imaginary time  $t \to t + i\beta$ is incompatible with the geometry of the global de Sitter manifold, the only exception being  $\beta = 2\pi$.

\section{Mode expansion of the holomorphic plane waves}
\label{4444}

Let us now move to the complex two-dimensional de Sitter spacetime:
\begin{equation}
    dS_2^c=\{Z\in {\bf C}^3,\ \ \  Z_0^2-Z_1^2-Z_2^2 = -1\}.
\end{equation}
We may use the same coordinate chart as in Eq. (\ref{coordinates}):
\begin{equation}
\label{coordinatesc}
 Z(t,x)=\begin{cases}
   Z^0=\sinh t \ \sech x\\
   Z^1=\tanh x \\
   Z^2= \cosh t \ \sech x
 \end{cases}.
\end{equation}
but now  $t$ and $x$ are complex. In particular
\begin{enumerate}
    \item
For $0< \Im\,  t <\pi$ and $x\in {\bf R}$ the point $Z(t,x)$ belongs to the forward tube
\begin{equation}
{\cal T}_+ = \{Z=X+iY \in dS_2^c,\ \ Y^2>0, \ \  Y^0>0\}.
\end{equation}
\item
For $-\pi < \Im\,  t <0$ and $x\in {\bf R}$ the point $Z(t,x)$ belongs to the backward tube
\begin{equation}
{\cal T}_- = \{Z=X+iY \in dS_2^c,\ \ Y^2>0, \ \  Y^0<0\}.
\end{equation}
\end{enumerate}
There exists
a remarkable set of solutions of the de Sitter Klein Gordon equation which
may be interpreted as de Sitter plane waves \cite{bm,bgm,gelfand}.
Their definition makes no appeal to any particular coordinate system  and may be given just in terms of the ambient spacetime coordinates:
given a forward pointing lightlike real vector $\xi$  in the ambient spacetime\footnote{ $\xi$ is a real vector belonging to the forward lightcone  $C^+= \{\xi \in {\bf R}^3,\ \ (\xi^0)^2- (\xi^1)^2-  (\xi^2)^2=0, \ \  \xi^0>0\}$.}
and a complex number $\lambda \in C$ let us construct the homogeneous function
\begin{equation}
Z\in dS_2^c \ : \ \ \  {Z}\mapsto (\xi\cdot Z)^\lambda.
\label{pwholo}
\end{equation}
For any given $\xi$ and $\lambda$ the above functions are  holomorphic in the
tuboids  $ {\cal T}_\pm$ \cite{bm,bgm}
and  satisfy the massive (complex) de Sitter Klein-Gordon equation:
\begin{equation}
(\Box-\lambda(\lambda-1))(\xi\cdot Z)^\lambda =0 \label{cdskg}
\end{equation}
(we may write $\lambda = -\frac 12 +i\nu$; in the following we will take for simplicity $\nu \in {\bf R}$). The boundary values
\begin{equation}
(\xi\cdot X)_\pm^\lambda = \lim_{Z\in {\cal T}_\pm, \, Z \rightarrow X}
(\xi\cdot Z)^\lambda
\label{p.1}
\end{equation}
are homogeneous distributions of degree $\lambda$ in the ambient spacetime
and their restrictions to the real  manifold $dS_2$
are solutions of the real de Sitter Klein-Gordon equation.
All these
objects are entire functions of  $\lambda$.

Let us now expand the above plane wave into modes of the static patch.
The first thing to be done is to choose a basis manifold of the forward light-cone; the convenient choice is the hyperbolic basis $\Gamma = \Gamma_l\cup\Gamma_r$  "parallel" to the coordinate system (\ref{coordinates}) of the static chart:
\begin{equation}
 \xi_l(w) = \left\{\begin{array}{lll}  \xi^0 &=& \cosh w  \cr
\xi^1&=&   - 1  \cr
\xi^2&=&  \sinh w
\end{array}
\right.
\ \ \ \
 \xi_r(w) = \left\{\begin{array}{lll}  \xi^0 &=& \cosh w  \cr
\xi^1&=&   + 1  \cr
\xi^2&=&  -\sinh w
\end{array}
\right. .
\label{coor}
\end{equation}
With all the above specifications, we get
\begin{equation}
\xi_l \cdot Z = \tanh x+ \sech x \sinh (t-w), \ \ \ \ \xi_r \cdot Z = -\tanh x+ \sech x \sinh (t+w).
\end{equation}
Let us take  $Z(t+i\epsilon,x)$ with $t$ real.
Since $Z(t+i\epsilon,x) \in {\cal T_+}$ the wave $(\xi_l \cdot Z)^\lambda$ is a regular function of $t$ decreasing at infinity; its Fourier transform is given by
\begin{equation} \label{20}
\int_{-\infty}^\infty e^{-i \omega t} \left(\xi_l(w) \cdot Z(t+i\epsilon,x )\right)^{-\frac 12 + i \nu}d t =     \frac {2  e^{-i\omega w }\Gamma(\frac 12 -i\nu  + i\omega)}{\Gamma(\frac 12 - i\nu )} e^{ \frac 12 \pi \omega}  \left( e^{  -\pi \omega}Q^{-i \omega}_{-\frac12 -i \nu}(\tanhx+i \epsilon)\right) ;
\end{equation}
here $Q$ is the associated  Legendre function of the second kind\footnote{Note that the above Legendre functions are related by complex conjugation as follows:
\begin{eqnarray}
 \left({Q^{-i\omega}_{-\frac 12 -i \nu}(u+i\epsilon) }\right)^*
=e^{2\pi\omega}{Q^{i\omega}_{-\frac 12 +i \nu}(u-i\epsilon) }. \label{cc}
\end{eqnarray}} \cite{Bateman} defined on the complex plane cut on the real axis from  $-\infty$ to 1.  Inversion gives
\begin{equation}
 (\xi_l(w) \cdot Z(t+i \epsilon,x) )^{-\frac 12 +i\nu}
  =\frac 1 {\pi}\int_{-\infty}^\infty  e^{i\omega(t-w)}  \frac {\Gamma(\frac 12 -i\nu  + i\omega)}{\Gamma(\frac 12 - i\nu )} e^{ \frac 12 \pi \omega}  \left( e^{  -\pi \omega}Q^{-i \omega}_{-\frac12 -i \nu}(\tanhx+i \epsilon)\right) d\omega.  \label{11}
\end{equation}
For $Z\in {\cal T}^-$  an analogous  computation gives
\begin{equation}
(\xi_l(w) \cdot Z(t-i \epsilon,x) )^{-\frac 12 -i\nu}
 = \frac 1 {\pi} \int e^{-i\omega(t-w)}  \frac {\Gamma(\frac 12 +i\nu  - i\omega)}{\Gamma(\frac 12 + i\nu )} e^{ \frac 12 \pi \omega}  \left( e^{  \pi \omega}Q^{i \omega}_{-\frac12 +i \nu}(\tanhx-i \epsilon)\right) d\omega. \label{22}
\end{equation}
Similarly
\begin{equation}
(\xi_r(w) \cdot Z(t\pm i \epsilon,x) )^{-\frac 12 \pm i\nu}=  \pm  \frac {  e^{-\nu\pi }}{i\pi} \int e^{\pm i\omega(t+w)}  \frac {\Gamma(\frac 12 \mp i\nu  \pm i\omega)}{\Gamma(\frac 12 \mp i\nu )} e^{ \frac 12 \pi \omega}  \left( e^{  \mp \pi \omega}Q^{\mp i \omega}_{-\frac12 \mp i \nu}(\tanhx\mp i \epsilon)\right) d\omega.
\label{33}
%(\xi_r(w) \cdot Z(t-i \epsilon,x) )^{-\frac 12 -i\nu} = \frac { i e^{-\nu\pi }}{\pi}  \int e^{-i\omega(t+w)}  \frac {\Gamma(\frac 12 +i\nu  - i\omega)}{\Gamma(\frac 12 + i\nu )} e^{ \frac 12 \pi \omega}  \left( e^{  \pi \omega}Q^{i \omega}_{-\frac12 +i \nu}(\tanhx+i \epsilon)\right) d\omega. \cr \label{44}
\end{equation}

\section{Mode expansion of the  maximally analytic two-point function}
\label{bd8}

The maximally analytic (Bunch-Davies) two-point function admits the following {\em global} manifestly de Sitter invariant integral representation, valid for $Z_1\in {\cal T}^-$ and $Z_2 \in \cal T^+$ \cite{bm,bgm}:
\begin{eqnarray}
\label{5.1}
W_{BD}(Z_1,Z_2)&=&  \frac {e^{\pi \nu}}{8\pi \cosh\pi \nu }  \int_{\Sigma}(\xi \cdot Z_1 )^{-\frac 12 -i\nu} (\xi \cdot Z_2 )^{-\frac 12 +i\nu} d\sigma(\xi).
\end{eqnarray}
Here $\Sigma$ is any basis manifold of the  forward lightcone $C_+$  and  $d\sigma$ the corresponding induced measure \cite{bm}.  In the symbol $W_{BD}$ referring to the Bunch-Davies Wightman function we left the mass parameter is $m=\sqrt{ \frac 14+\nu^2}$ implicit.

By using the static coordinates (\ref{coordinates}), the basis $\Gamma = \Gamma_l\cup\Gamma_r$ for the lightcone (with $d\sigma_{\Gamma_i} = dw$) and by inserting Eqs. (\ref{11} -- \ref{33})  in Eq. (\ref{5.1}) we get that the boundary value on the reals {\em in the static chart} of the above {\em global} holomorphic two-point function can be represented as follows:
\begin{eqnarray}
\label{5.2}
W_{BD}(X_1,X_2) &=&   \int_{-\infty}^\infty   {e^{-i\omega(t_1-t_2)}} \widetilde W_{BD}(\omega, u_1,u_2)  d\omega
 \\ && \cr
\widetilde W_{BD}(\omega, u_1,u_2) &= &   \frac {e^{\pi \omega}} {4 \pi^2 \cosh\pi(\nu-\omega)} \left[e^{\pi\nu} Q^{i \omega}_{-\frac12 +i \nu}(\tanhx_1-i \epsilon) Q^{-i \omega}_{-\frac12 -i \nu}(\tanhx_2+i \epsilon)\right. + \,   \cr &+&
 \left. e^{-\pi\nu} Q^{i \omega}_{-\frac12 +i \nu}(\tanhx_1+i \epsilon) Q^{-i \omega}_{-\frac12 -i \nu}(\tanhx_2-i \epsilon) \right]\label{bbdd}
 \label{bbdd0}
\end{eqnarray}
By using the  identity \cite{Bateman}
\begin{align}
\label{3}
Q^{i\omega}_{-\frac{1}{2}+i\nu}(u \pm i0)=\frac{\pi}{2\cosh(\pi(\nu+\omega))}e^{-\pi\omega\mp \frac{\pi\omega}{2}}\left(\mp i e^{\pm\pi(\nu+\omega)}\mathsf{P}^{i\omega}_{-\frac{1}{2}+i\nu}(u)+\mathsf{P}^{i\omega}_{-\frac{1}{2}+i\nu}(-u) \right)
\end{align}
a  straightforward calculation  shows that
\begin{equation}
  \widetilde W_{BD}(\omega, u_1,u_2)=  \widetilde P_{\nu}(\omega,u_1,u_2).
\end{equation}
When $\beta=2\pi$ Eqs.  \eqref{wightman} and (\ref{bbdd})  do coincide  proving the claimed identification.
\vskip 20 pt

The permuted two-point function is in turn represented as follows:
 \begin{equation}
 W_{BD}(X_2,X_1)
=
\int_{-\infty}^\infty   {e^{-i\omega(t_1-t_2)}}  \widetilde P_\nu(-\omega, u_2,u_1)  d\omega =
 \int_{-\infty}^\infty   {e^{-i\omega(t_1-t_2)}} e^{-2\pi\omega} \widetilde P_\nu(\omega, u_1,u_2)  d\omega .
\label{kkms}
\end{equation}
In the above chain of identities we changed the integration variable $\omega \to -\omega$ and -  in the second step -  used the symmetry of the two-point function $\nu\to-\nu$.
As a by product - by the Riemann-Lebesgue theorem - we get also the following  crucial  identity (which may also be checked directly):
\begin{eqnarray}
\widetilde P_\nu(-\omega, u_2,u_1) = e^{-2\pi\omega}\widetilde P_\nu(\omega, u_1,u_2).
 \label{rel1}
\end{eqnarray}
Eq. (\ref{kkms}) encodes the {Kubo-Martin-Schwinger} property of the restriction of the maximal analytic two-point function (\ref{bbdd}) to the static patch:  a geodetic observer in the static patch "perceives" a  thermal bath of particles at inverse temperature $2\pi R$.

\section{More about the vacuum of the static geodetic observer}

By using Eqs. (\ref{5.2}) and (\ref{kkms})
we obtain the following new integral representation of the covariant commutator in the static chart:
 \begin{eqnarray}
 C_\nu(X_1,X_2) = W_{BD}(X_1,X_2)- W_{BD}(X_2,X_1) =   \int_{-\infty}^\infty   {e^{-i\omega(t_1-t_2)}}  \widetilde C_\nu(\omega, u_1,u_2)  d\omega \label{comm}
\end{eqnarray}
where
%\footnote{There is no subscript $\nu$ in the commutator because it does not depend on the state we used to compute it (the notation $W_\nu$ referring only to the BD state).  Of course the commutator depends on the mass parameter.}
\begin{eqnarray}
 \widetilde{C}_\nu(\omega, u_1,u_2) =  (1-e^{-2\pi\omega})\, \widetilde P_\nu(\omega, u_1,u_2)
= -  \widetilde C_\nu(-\omega, u_2,x_1) \label{rel1b}
\end{eqnarray}
Let us take the zero temperature  limit $\beta \to \infty$ in Eq. (\ref{wightman}); only positive energies survive:
\begin{eqnarray}
W_\infty(X_1,X_2) =   \int_{0}^\infty   {e^{-i\omega(t_1-t_2)}} (1-e^{-2\pi\omega})\, \widetilde P_\nu(\omega, u_1,u_2)\, d\omega \cr = \int_{-\infty}^\infty   {e^{-i\omega(t_1-t_2)}} \theta(\omega)\, \widetilde C_\nu(\omega, u_1,u_2)\, d\omega. \label{vacuum0}
\end{eqnarray}
 $\theta(\omega)$ is Heaviside's step function. The above equation points towards the  following natural family of  Rindler-de Sitter positive frequency modes ($\omega\geq 0$):
 \begin{eqnarray}
\varphi_{\omega,1}(t,u)  &=&  e^{\frac 12 \pi \nu}  e^{\pi\omega} \sqrt{ \frac { \sinh{\pi\omega} }{ {2\pi^3}}}\  \Gamma\left(\frac 12 - i\nu + i \omega \right) \  {e^{-i\omega t }}
Q^{i \omega}_{-\frac12 +i \nu}(u-i \epsilon)
\cr
\varphi_{\omega,2}(t,u)  &=&  e^{-\frac 12 \pi \nu}  e^{\pi\omega}   \sqrt{ \frac { \sinh{\pi\omega} }{ {2\pi^3}}}\  \Gamma\left(\frac 12 - i\nu + i \omega \right)\ {e^{-i\omega t }}
Q^{i \omega}_{-\frac12 +i \nu}(u + i \epsilon),
\cr
\varphi^*_{\omega,1}(t,u)  &=&  e^{\frac 12 \pi \nu}  e^{-\pi\omega} \sqrt{ \frac { \sinh{\pi\omega} }{ {2\pi^3}}}\  \Gamma\left(\frac 12 + i\nu - i \omega \right) \  {e^{i\omega t }}
Q^{-i \omega}_{-\frac12 -i \nu}(u+i \epsilon)
\cr
\varphi^*_{\omega,2}(t,u)  &=&  e^{-\frac 12 \pi \nu}  e^{-\pi\omega}   \sqrt{ \frac { \sinh{\pi\omega} }{ {2\pi^3}}}\  \Gamma\left(\frac 12 + i\nu - i \omega \right)\ {e^{i\omega t }}
Q^{-i \omega}_{-\frac12 -i \nu}(u {-} i \epsilon). \label{modes}
\end{eqnarray}
equivalent to the one used in Sect. \ref{quantize}.
%is complete in the Klein-Gordon sense, where the normalization surface is the half-circle $\{X^{\alpha}X_\alpha = -1, X^0=0, X^2>0\}$ (or any surface homotopical to that half-circle).
Using the above modes we may  represent the field operator in the static patch in the  usual way
\begin{equation}
\phi(t,x) = \int_0^\infty  \left(\varphi_{\omega,1}(t,u) a_1(\omega) +\varphi_{\omega,2}(t,u) a_2(\omega) + \varphi^*_{\omega,1}(t,u) a^\dagger_1(\omega) +\varphi^*_{\omega,2}(t,u) a^\dagger_2(\omega) \right) d\omega.
\end{equation}
The state $W_\infty(X_1,X_2)$ is characterized by the conditions
$
 a_1(\omega)  \Psi_0 =      a_2(\omega)  \Psi_0 = 0 ;
$
it is a {\em pure state}
\begin{eqnarray}
W_\infty(X_1,X_2)=
 \sum_i \int_0^\infty \varphi_{\omega,i}(t_1,u_1)  \varphi_{\omega,i}^* (t_2,u_2) d\omega. \label{pure}
 \end{eqnarray}
Positive-definiteness is also clear from (\ref{pure}). The state defined by  $W_\infty$ may be interpreted as the vacuum state for the geodesic observer in the Rindler-de Sitter wedge and is the close analogous of the Fulling vacuum of Rindler QFT \cite{ful1,ful2}.

Finally,  the covariant commutator (\ref{comm}), which is of course independent from the chosen state, can be written as follows:
 \begin{eqnarray}
&& C_\nu(X_1,X_2) = W_\infty(X_1,X_2) - W_\infty(X_2,X_1) = \cr
&&= \sum_{i=1}^2 \int_0^\infty [ \varphi_{\omega,i}(t_1,u_1)  \varphi_{\omega,i}^* (t_2,u_2) -  \varphi_{\omega,i}(t_2,u_2)  \varphi_{\omega,i}^* (t_1,u_1)] d\omega.
 \label{compl}
\end{eqnarray}

 \section{Other time-translation invariant states.}

 More generally,  we may introduce the two-point functions
 \begin{eqnarray}
 W_F(X_1,X_2) &=&  \int_{-\infty}^\infty   {e^{-i\omega(t_1-t_2)}} F(\omega) \widetilde C_\nu(\omega, u_1,u_2)  \,d\omega
\\
 W_F(X_2,X_1) &=&
 \int_{-\infty}^\infty   {e^{-i\omega(t_1-t_2)}} F(-\omega) \widetilde C_\nu(-\omega, u_2,u_1)  \, d\omega
\end{eqnarray}
where $F(\omega)$ is a real function or a distribution such that the product $F(\omega) \widetilde C(\omega, u_1,u_2) $ is well defined. Eq. (\ref{rel1b}) implies that it must be
\begin{eqnarray}
F(\omega)+ F(-\omega) =1.
\end{eqnarray}
In particular
\begin{enumerate}
    \item The vacuum (\ref{vacuum0}) of the static geodetic observer correspond to
    \begin{equation}
        F(\omega) = \theta(\omega).
         \end{equation}
\item The Bunch-Davies maximally analytic state  (\ref{bbdd0}) correspond to
    \begin{equation}
        F(\omega) = \frac{1}{1-e^{-2\pi\omega}}.
         \end{equation}
\item    An antisymmetric function $\beta(-\omega) = -\beta(\omega)$ defines a time invariant state
     \begin{equation}
        F(\omega) = \frac{1}{1-e^{-\beta(\omega)}}.   \end{equation}
 \item     The thermal equilibrium state (\ref{wightman}) at inverse temperature $\beta$ corresponds to $\beta(\omega) = \beta\, \omega$
     %\begin{equation}  F(\omega) = \frac{1}{1-e^{-\beta\omega}}.   \end{equation}
        \end{enumerate}
%Any of the above states defines a local canonical quantum field  in the static patch satisfying the canonical commutation relations and positive-definite (provided the integral converges in the sense of distributions).

All the above two-point functions have the following general structure
\begin{eqnarray}
 W (X_1,X_2) =
 \sum_{i=1,2} \int_0^\infty\left. \cosh^2 (\gamma(\omega)) \ \varphi_{\omega,i}(t_1,u_1) \  \varphi_{\omega,i}^* (t_2,u_2) \right. d\omega\,  + \cr
 +\sum_{i=1,2} \int_0^\infty  \sinh^2 (\gamma (\omega)) \ \varphi^*_{\omega,i}(t_1,u_1)  \varphi_{\omega,i} (t_2,u_2)  \, d\omega \label{bog}
 \end{eqnarray}
with in particular
$
 \cosh \gamma(\omega) =  {(1-e^{-\beta \omega})^{-\frac 12} } \label{jjj}
$
for the thermal state

\begin{eqnarray}
W_\beta (X_1,X_2) =
 \sum_{i=1,2} \int_0^\infty  \frac{\varphi_{\omega,i}(t_1,u_1) \  \varphi_{\omega,i}^* (t_2,u_2)}{1-e^{-\beta \omega }}  \  d\omega\,
 +\sum_{i=1,2} \int_0^\infty \frac{\varphi^*_{\omega,i}(t_1,u_1)  \varphi_{\omega,i} (t_2,u_2)}{e^{\beta \omega }-1} \   d\omega. \label{wigbeta}
 \end{eqnarray}
  The latter formula in turn  allows to write $W_{\beta} (t, x_1, x_2)$ as a Matsubara sum over imaginary frequencies as follows:
\begin{eqnarray} W_{\beta} (X_1,X_2)& = &
 \sum_{n=0} ^\infty W_\infty(t_1-in \beta,x_1, t_2, x_2) +
 \sum_{n=1} ^\infty W_\infty(t_2,x_2, t_1+in \beta, x_1 ). \label{matsubara}
  \end{eqnarray}
The above representations clearly shows that all such states (but the vacuum $\gamma = 0$) are mixed states. In particular the maximally analytic Bunch-Davies two-point function is written
\begin{equation} W_{BD} (X_1,X_2) =
 \sum_{i=1,2} \int_0^\infty  \frac{\varphi_{\omega,i}(t_1,u_1) \  \varphi_{\omega,i}^* (t_2,u_2)}{1-e^{-2\pi \omega }}  \  d\omega\,
 +\sum_{i=1,2} \int_0^\infty \frac{\varphi^*_{\omega,i}(t_1,u_1)  \varphi_{\omega,i} (t_2,u_2)}{e^{2\pi \omega }-1}    d\omega.  \label{bd2}
 \end{equation}
These are  simple  examples of  what has been called a "Generalized Bogoliubov transformation" \cite{schaeffer,sch2}, a construction that directly provides mixed states by suitably extending the canonical quantization formalism.

\section{Alpha--states in the static chart}

The set of states (\ref{jjj})  does not contain every time translation invariant state. There is still the freedom to add to the  two-point function a symmetric part that, as such, does not contribute to the commutator. The so-called $\alpha$-vacua \cite{ss,Akhmedov:2019esv, Mottola,Allen,Epstein} belong to this second class of states. Let us briefly sketch their construction in the static patch coordinates.

The two-point Wightman functions of the $\alpha$-vacua may be written in terms of the Bunch-Davies two-point function as follows \cite{Allen,Epstein}:
\begin{eqnarray}
W^{(\alpha)} (X_1,X_2)
= \cosh^2 \alpha \ W_{BD}(X_1,X_2) + \sinh^2 \alpha \ W_{BD}(X_2, X_1) +   \cr
+ \frac 12  \sinh 2\alpha \ [W_{BD}(X_1,- X_2)
+ W_{BD}(-X_1,\  X_2)]
\label{tpalpha}
\end{eqnarray}
We are left with the task of expanding  $W_{BD}(X_1,- X_2) $
in the modes (\ref{modes}).
To do it, let us introduce the parity automorphism  of the static patch :
\begin{equation}
 X(t,x)\rightarrow \widetilde X(t,x) =  X(t,-x)
 %= \left\{\begin{array}{lll}  X^0 &=& \sinh t \sin \theta  \crX^1&=&  -\cos \theta  \crX^2&=& \cosh t \sin \theta\end{array}\right.\
\end{equation}
The curve
$
s\rightarrow  \widetilde X(t+ is,x) $ for $0<s<\pi$ is entirely contained in ${\cal T}_+$ and ends at
\begin{equation}
\widetilde X(t+ i\pi,x) = - X(t,x)
\end{equation}
in the left Rindler--de Sitter wedge (see Fig. (\ref{fig:spatch})).
Similarly, the curve
$
s\rightarrow  \widetilde X(t+ is,x) $ for $0>s>-\pi$ is entirely contained in ${\cal T}_-$ and ends again at
$
\widetilde X(t- i\pi,x) = - X(t,x)
$ but from the opposite tube.

Given any two points $X_1$ and $X_2$ in the right  Rindler--de Sitter wedge  we may
 use again  the maximally analytic {\em global} two-point function (\ref{5.1}) and get
\begin{eqnarray}
W_{BD}(X_1,-X_2)&=&  W_{BD}(X_1,\tilde X_2(t_2+i \pi , x_2))
= W_{BD}(-X_1,X_2) =   W_{BD}(\tilde X_1(t_1-i \pi , x_1),X_2)
 \cr
  \cr
&=&  - \frac {i} {4\pi^2} \int_{-\infty}^\infty   \frac {e^{-i\omega(t_1-t_2)}}{\cosh\pi(\nu-\omega)} Q^{i \omega}_{-\frac12 +i \nu}(u_1-i \epsilon) Q^{-i \omega}_{-\frac12 -i \nu}(u_2-i \epsilon) d\omega \,   \cr
 &+&  \frac {i} {4\pi^2} \int_{-\infty}^\infty    \frac {e^{-i\omega(t_1-t_2)}}{\cosh\pi(\nu-\omega)}  Q^{i \omega}_{-\frac12 +i \nu}( u_1+i \epsilon) Q^{-i \omega}_{-\frac12 -i \nu}(u_2+i \epsilon) d\omega
\\
\cr
\cr &=& -i \int_0^\infty \frac{ \varphi_{\omega,1}(t_1,x_1) \  \varphi_{\omega,2}^* (t_2,x_2) -   \varphi_{\omega,2}(t_1,x_1) \  \varphi_{\omega,1}^* (t_2,x_2)}{2 \sinh{\pi \omega }}\ d\omega
\cr &+&  i \int_0^\infty \frac{ \varphi^*_{\omega,1}(t_1,
u_1)  \varphi_{\omega,2} (t_2,u_2) -  \varphi^*_{\omega,2}(t_1,u_1)  \varphi_{\omega,1} (t_2,u_2)
 } {2 \sinh \pi \omega} \ {d\omega}. \label{alpha0}\end{eqnarray}
In the second step we used Eq. (\ref{bd2}) and the following relations:
 \begin{eqnarray}
&&\varphi_{\omega,1}(t\pm i\pi,-u)  =  i  e^{\pm \pi\omega}   \varphi_{\omega,2}(t,u),
\ \ \ \ \
\varphi_{\omega,2}(t\pm i \pi,-u)
= - i e^{\pm \pi\omega}   \varphi_{\omega,1}(t,u),
\cr
&&\varphi^*_{\omega,1}(t\pm i \pi,-u)   =   -i e^{\mp \pi\omega}   \varphi^*_{\omega,2}(t,u),
\ \ \
\varphi^*_{\omega,2}(t\pm i \pi,-u)   =   i e^{\mp \pi\omega}   \varphi^*_{\omega,1}(t,u).
\end{eqnarray}
Integration is (\ref{alpha0})  the over positive energies only. Putting everything together we get:

\begin{eqnarray} W^{(\alpha)} (X_1,X_2) &= &
 \sum_i \int_0^\infty  \cosh^2(\gamma(\omega))\,  {\varphi_{\omega,i}(t_1,u_1) \  \varphi_{\omega,i}^* (t_2,u_2)}  \  d\omega\, \cr
 &+&\sum_i \int_0^\infty  \sinh^2(\gamma(\omega))\,   {\varphi^*_{\omega,i}(t_1,u_1) \,  \varphi_{\omega,i} (t_2,u_2)} \   d\omega
 \cr  &-& i \sinh 2\alpha \int_0^\infty \frac{ \varphi_{\omega,1}(t_1,x_1) \  \varphi_{\omega,2}^* (t_2,x_2) -   \varphi_{\omega,2}(t_1,x_1) \  \varphi_{\omega,1}^* (t_2,x_2)}{2 \sinh{\pi \omega }}\ d\omega
\cr &+&  i\sinh{2\alpha} \int_0^\infty \frac{ \varphi^*_{\omega,1}(t_1,u_1)  \varphi_{\omega,2} (t_2,u_2) -   \varphi^*_{\omega,2}(t_1,u_1)  \varphi_{\omega,1} (t_2,u_2)
 } {2 \sinh \pi \omega} \ {d\omega} \label{alpha10}\end{eqnarray}
where
\begin{equation}
    \cosh(\gamma(\omega)) = \sqrt{ \frac{e^{\pi\omega}\cosh^2\alpha +e^{-\pi\omega}\sinh^2\alpha }{2\sinh(\pi \omega)}}.
\end{equation}
As expected, the $\alpha$-vacua are  translation invariant w.r.t. the  time variable  of the Rindler - de Sitter wedge; here the generalized Bogoliubov transformation of the positive energy modes is  more general than the one exhibited in Eq. (\ref{bog}).
The extra terms which do not contribute to the commutator are altogether  symmetric in the exchange of $X_1$ and $X_2$.

\section{More about thermal propagators}
\label{wight}

In this section we examine some properties of the thermal correlation functions and discuss various limiting behaviors. This study is to better characterize them and also to lay the ground for the study of the IR loop contributions which will be the matter of a companion paper.

\subsection{Wightman propagators for large time--like separation}

Let us consider the  limit $t = t_1 - t_2 \rightarrow{\infty}$, $x_1=x_2=0$ (the general case $x_1 \not= x_2$ being essentially the same).  The integrand in Eq. \eqref{wightman}  has  poles at
\begin{align}
	\label{polb}
\omega =\pm \,\mu - \frac{i}{2} + i n, \quad n \in \mathbb{Z}, \quad \quad \omega=\frac{2\pi i k}{\beta}, \quad  k \in \mathbb{Z},\, \, k \neq 0 .
\end{align}
In Eq. \eqref{wightman} there is no pole at $\omega = 0$; still, $\omega= 0$  has a role to play in calculating the spacelike asymptotics.

In the limit $t\to \infty$ the leading contributions come from the poles %in Eq. \eqref{wightman}
which are closer to the real axis:
\begin{equation}
\label{beh}
W_\beta(t, x_1=x_2=0)\approx
 \begin{cases}
   e^{-\frac t 2 }(C_+ e^{ i\mu t} + C_- e^{- i\mu t }) & \text{for} \ \  \beta < 4\pi \\
   C_\beta e^{-t \frac{2\pi}{\beta}} & \text{for} \ \ \beta > 4\pi
   \end{cases},
\end{equation}
where
%\begin{align}
%C_\beta=-\frac{\pi}{\beta}\frac{\sin\frac{2\pi^2}{\beta}}{\cosh\pi(\frac{2\pi i}{\beta}-\mu)\cosh\pi(\frac{2\pi i}{\beta}+\mu)} \mathsf{P}^{\frac{2\pi}{\beta}}_{-\frac{1}{2}+i\mu}(0)\mathsf{P}^{-\frac{2\pi }{\beta}}_{-\frac{1}{2}+i\mu}(0).
%\end{align}
\begin{align}
C_+=  C_-^*=\frac{1-e^{-2\pi(\nu+\frac{i}{2})}}{1-e^{-\beta(\nu+\frac{i}{2})}}\frac{e^{-\pi\mu}\Gamma\Big( i\mu \Big)\Gamma\Big(\frac{1}{2}-i\mu\Big)}{2\pi \sqrt{\pi}},
\\
    C_{\beta}=\frac{\sin\left(\frac{2\pi^2}{\beta}\right)}{4\pi^2\beta}  \Big|\Gamma\Big(\frac{1}{4}-\frac{i\mu}{2}-\frac{\pi}{\beta}\Big) \Gamma\Big(\frac{1}{4}-\frac{i\mu}{2}+\frac{\pi}{\beta}\Big) \Big|^2.
\end{align}
%
%The de Sitter invariant case $\beta = 2\pi$  is a special instance of Eq. (\ref{}) this expansion corresponds to the large $\zeta \approx e^\frac{t}{2}$ behavior of the Wightman propagator, which is well known from the large $\zeta$ expansion of (\ref{2DBDpropagator}).
The asymptotic behavior of the propagator changes at $\beta=4\pi$.
In the limit $\beta \to \infty$ the constant $C_\beta $ tends to zero and the Wightman function asymptotics is given again the upper line in (\ref{beh}).

The first set of poles in (\ref{polb}) %$\omega=(\pm \mu - \frac{i}{2} + i n)$
 is also actually related to the transmission and reflection coefficients of the quantum mechanical scattering problem (\ref{QM}). For instance
%\begin{align} \label{asympg} \psi_{\omega}(u)\approx \sqrt{\sinh(\pi \omega)}\Gamma(-i\omega)\begin{cases}e^{i\omega x}+A_-(\omega)e^{-i\omega x} \quad \text{as}\ \  x\to-\infty\\ A_+(\omega)e^{i\omega x} \quad \text{as}\ \ x\to \infty \end{cases}\end{align}
%\begin{align}   A_-(\omega) =  \frac{\cosh({\pi \nu}) \, \Gamma(i\omega) \, \Gamma\left(\frac12+i\mu-i\omega\right) \, \Gamma\left(\frac12-i\mu-i\omega\right) }{\pi \, \Gamma(-i\omega) }  ,\\    A_+(\omega) =  \frac{\Gamma\left(\frac12+i\mu-i\omega\right) \, \Gamma\left(\frac12-i\mu-i\omega\right)  }{ \Gamma(1-i\omega)\, \Gamma(-i\omega)} .
%\end{align}
a straightforward computation based on Eqs. (\ref{plus}) and (\ref{minus}) gives the transmission coefficient
\begin{align}
    T= \frac{\sinh^2(\pi\omega)}{\cosh[\pi (\mu-\omega)] \, \cosh[\pi (\mu+\omega)\, ]}.
\end{align}
%\begin{align*}
%    \omega=\pm \mu +\frac{i}{2}+i\pi n,
%\end{align*}
%which is the second chain of poles from \eqref{polb}, because the transmission and reflection coefficients contribute to the integrand in the Wightman propagator in Eq. \eqref{wightman} or else  Eq. \eqref{bog}.
The second set of poles in \eqref{polb} depends also on the inverse temperature $\beta$. %, and comes from the  thermal distribution in the integral representation of the Wightman propagator.
As  $\beta$ increases the poles move towards the real axis.
%When $\beta < 4\pi$  poles of the first set closer to the real axis.
When $\beta > 4\pi$ poles of the second set dominate and the large $t$ behavior of the propagator changes accordingly.

\subsection{Large space--like separation}
\label{horrr}

We now  evaluate  the asymptotic behaviour of the correlators  when one of the spacelike coordinates goes to infinity in two distinct ways.

By use of the asymptotic behaviour of the Ferrers function at $x\to \infty$ we get
\begin{eqnarray}
\label{plus}
 && \mathsf{P}^{i\omega}_\ind\left(\tanh x\right) \underset{x\to\infty}{\approx} \frac{e^{i\omega x}}{\Gamma(1-i\omega)},
\\
\label{minus}
&& \mathsf{P}^{i\omega}_\ind\left(-\tanh x\right) \underset{x\to\infty}{\approx}  \bigg[ \frac{\Gamma\big(-i\omega\big) e^{-i\omega x}}{\Gamma\big(\frac12+i\mu-i\omega\big)\Gamma\big(\frac12-i\mu-i\omega\big)}+\frac{\cosh(\mu\pi)\Gamma\big(i\omega\big) \, e^{i\omega x}}{\pi }\bigg] .
\end{eqnarray}
The singularities at $\omega = 0$ in the latter equation cancel each other, but, in the limit under consideration the two terms  contribute separately.
%because according to Jordan's lemma we will have to close contours for them in different halves of the complex $\omega$--plane. Hence, although Eq. \eqref{wightman} does not have a pole in $\omega = 0$ it will come from one of the terms in (\ref{minus}).
By substituting the above  expressions into \eqref{wightman}  and making the shift $\omega \to \omega+ i \epsilon$  %the contour in the lower half plane for the $\sim e^{-i\omega x_2}$ part and in the upper half plane for $\sim e^{i\omega x_2}$ part of the integrand in:
%\begin{gather}	W_\beta(t,x_1,x_2\to + \infty) 	\approx 	\int\limits^\infty_{-\infty} \frac{d\omega}{4 } \frac{e^{i\omega t}}{e^{\beta(\omega+i0)}-1} \frac{\sinh \pi \omega}{ \cosh\pi\left(\omega-\mu\right)\cosh\pi\left(\omega+\mu\right)} \Bigg[ \mathsf{P}^{i\omega}_\ind\left(\tanh x_1\right)  \frac{e^{-i\omega x_2}}{\Gamma(1+i\omega)}+ \notag\\ +	\mathsf{P}^{i\omega}_\ind\left(-\tanh x_1\right)\bigg[ \frac{\Gamma\big(i\omega\big) e^{i\omega x_2}}{\Gamma\big(\frac12+i\mu+i\omega\big)\Gamma\big(\frac12-i\mu+i\omega\big)}+\frac{\cosh(\mu\pi)\Gamma\big(-i\omega\big) \ e^{-i\omega x_2}}{\pi }\bigg]\Bigg]  	.\end{gather}
%The limit under consideration is infrared. Hence, it is not surprising that
we see that the dominant contribution comes from the lower half plane. We get that in the limit $x_1 \to \infty$  the Wightman propagator still depends on the temperature:
\begin{align}\label{edgebeh}
\lim_{x_2\to \infty} 	W_{\beta}(t_1-t_2, x_1,x_2) = \frac{2\pi}{\beta}\frac{1}{4\cosh \mu \pi} P_\ind\left(-\tanh x_1\right) = \frac{2\pi}{\beta} W_{BD}(-\tanh x_1).
\end{align}

\vskip 10pt
Alternatively, we may consider the  formal Taylor expansion of Eq. (\ref{wightman}):
 % \begin{eqnarray}\frac{1-e^{-2\pi\omega}}{1-e^{-\beta \omega}}= \frac{2 \pi }{\beta }+ \left(\pi -\frac{2 \pi ^2}{\beta }\right) \omega +\left(\frac{\pi  \beta }{6}+\frac{4 \pi ^3}{3 \beta }-\pi ^2\right) \omega ^2+\left(-\frac{\pi ^2 \beta }{6}-\frac{2 \pi ^4}{3 \beta }+\frac{2 \pi ^3}{3}\right) \omega ^3+\ldots  \end{eqnarray}
%By insering this expansion in Eq. (\ref{wigg}) we get
 \begin{eqnarray}
 \label{wigg2}
&&   W_\beta(t,\, x_1,x_2)
= \frac{2 \pi }{\beta }W_{BD}(\zeta)
+ \left(\pi -\frac{2 \pi ^2}{\beta }\right) \   i \,\frac {\partial}{\partial t}W_{BD}(\zeta)
\cr   &-& \left(\frac{\pi  \beta }{6}+\frac{4 \pi ^3}{3 \beta }-\pi ^2\right)   \ \frac {\partial^2}{\partial t^2}W_{BD}(\zeta)
- \left(-\frac{\pi ^2 \beta }{6}-\frac{2 \pi ^4}{3 \beta }+\frac{2 \pi ^3}{3}\right)  \  i  \,  \frac {\partial^3}{\partial t^3}W_{BD}(\zeta)
 +\ldots
 \end{eqnarray}
For $\beta = 2\pi$ all the de Sitter breaking terms  (i.e. every term but the first) at the RHS cancel, as expected.
Also, when $t=t_1-t_2$ is held constant and either $x_1$ or $x_2$ tend to plus or minus infinity, only the first terms at the RHS survives, with $\zeta= - \tanh x$.

\subsection{Light-like separation}

For light--like separations the propagators should behave as in  Minkowski space. In the Bunch-Davies invariant case % $\beta=2\pi$
%(\ref{Zdist}) to obtain:
%\begin{align*} \zeta = - \frac{\cosh(t)+\sinh(x_1)\sinh(x_2)}{\cosh(x_1)\cosh(x_2)}\approx - 1,\quad \text{hence} \quad  \cosh(t) \approx \cosh(x_2-x_1) \quad \text{and} \quad t\approx \pm (x_2-x_1).
%\end{align*}
this comes immediately from Eq.  (\ref{2DBDpropagator}):

\begin{align}
\label{pec}
W_{BD} (\zeta \approx - 1) \approx -\frac{1}{4\pi} \log(1+\zeta) \approx -\frac{1}{4\pi} \log \Big[t^2-(x_1-x_2)^2 \Big].
\end{align}
For arbitrary $\beta$ at light-like separation large values of  $\omega$'s dominate in the integral \eqref{wightman}. For large $\omega$ we may approximate
$
\mathsf{P}^{i\omega}_\ind\left(\tanh x_1\right)\approx {e^{i\omega x_1}}/{\Gamma(1-i\omega)}.% \quad \text{as} \ \ \omega \to \infty.
$
%A straightforward calculation gives for the
and get the leading term %with the logarithmic precision:
\begin{gather*}
	W_\beta(t, x_1, x_2) \approx  \int\limits_{1}^{\infty} \frac{d\omega}{2\pi} \frac{e^{-i\omega t}}{2\omega} \Big(e^{-i\omega (x_1-x_2)}+e^{i\omega (x_1-x_2)}\Big)\approx -\frac{1}{4\pi} \log\Big[t^2-(x_2-x_1)^2\Big].
	\end{gather*}
The cutoff in this integral is order of  $R$ --- the radius of the de Sitter universe, which we set equal to one. The approximation  works for $|\omega|$ much larger than $m$ and $R$. The   dependence on the temperature is lost in this high energy limit: only the Hadamard term survives. %We will see below that for $\beta \ne 2\pi$ there is an extra singularity at the horizon related to the phenomenon discussed in the previous subsection, {despite the fact that events on the horizon also have  a light--like separation}.

\subsection{Anomalous singularities at the horizon  }

%Apart $\beta=2\pi$ for arbitrary $\beta$ the propagator $ W_\beta(t, x_1, x_2)$  is not a function of the invariant $\eqref{Zdist}$ anymore  $W_\beta(t, x_1,x_2)\neq W(\zeta_{12})$. E.g. in \cite{Akhmedov:2019esv} it was shown that for $\beta=\infty$ the propagator does not respect the de Sitter isometry. Essentially that is because for $\beta\ne2\pi$ the Wightman propagator has a wrong period to be a function of the invariant $\zeta$. In fact, de Sitter isometry requires that the propagator should be periodic under the shift $t \to t+2\pi i$, while the thermal Wightman function with arbitrary $\beta \neq 2\pi$ is periodic under the shift $t \to t+\beta i$.

When the temperature is an integer multiple of the Hawking-Gibbons temperature, i.e.  when $\beta={2\pi}/{N}$, we may use Eq. (\ref{wightman}) to derive  another  representation of the two-point function as a finite sum of Legendre functions (as oppposed to the infinite Matsubara-type series (\ref{matsubara})); this is obtained by translating the Bunch--Davies maximal analytic two-point function in the imaginary time variable {within the   analyticity strip $(-2\pi < \Im t < 0 )$} (see also \cite{ Akhmedov:2019esv,bertola}):
\begin{eqnarray}
 \label{wightman00}
   W_{\frac {2\pi}{N}}(t_1-t_2,\, x_1,x_2)
&=& \int_{-\infty}^{+\infty} {e^{-i\omega (t_1-t_2)} }\frac{{1-e^{-2\pi\omega}}}{{1-e^{-\frac{2\pi \omega}{N}}}}\widetilde P_{\omega,\nu}(u_1,u_2)\ d\omega =
%\label{2pin}
\cr
&&\cr &&\cr
\label{horprop}
&=&  \frac{1}{4\cosh\pi\nu} P_\ind \bigg( \zeta\bigg(t_1-t_2-i\epsilon, x_1,x_2\bigg) \bigg)  +\cr &+& \frac{1}{4\cosh\pi\nu} \sum_{n=1}^{N-1} P_\ind \left(\zeta \left(t_1-t_2  -i \frac{ 2\pi n }{N}, x_1,x_2\right)\right) .
\end{eqnarray}
The  first term on the RHS is exactly the Bunch--Davies de Sitter invariant Wightman function; this is singular at $\zeta = -1$.
The extra terms become singular when the two points approach either the left or the right horizon: %(either both $x_1 $ and $ x_2$ go to minus infinity  or  both $x_1$ and $x_2$ go to plus infinity).
\begin{equation}
\label{horcor}
X_1= X(\lambda + c_1, \lambda), \ \ \ \
X_2= X(\lambda + c_2, \lambda + \Delta \lambda) %x_1=\lambda,\qquad x_2=\lambda+\Delta x,\qquad t_1=\lambda+c_1, \qquad t_2=\lambda+c_2.
\end{equation}
In the limit $\lambda \to \pm \infty$ the above events belong to the horizons. Then:

\begin{align*}
\zeta \left(c_1-c_2  -i \frac{ 2\pi n }{N}, \lambda,\lambda+\Delta \lambda \right)=-\frac{\cosh \left(c_1-c_2  -i \frac{ 2\pi n }{N}\right) + \sinh \lambda \sinh(\lambda+\Delta \lambda)}{\cosh \lambda \cosh(\lambda+\Delta \lambda)} \to - 1. %\quad \text{as}  \ \ \lambda \to \infty.  \xrightarrow[\lambda\to \infty]{} 1
\end{align*}
%It means that both terms in \eqref{horprop} become singular on the horizon --- on its right upper edge. The same is true when both points together approach any of the edges of the horizon.
For generic $\beta$, the limit $\lambda\to\infty$ may be obtained by performing
%one can see the same fact from the mode expansion of $W_\beta(t, x_1, x_2)$. In fact, performing
manipulations similar to those which led to  \eqref{edgebeh}: %for the case \eqref{horcor}
 %\begin{align*}W_\beta(t,x_1,x_2) \approx \int d\omega\frac{e^{i\omega (c_2-c_1)}}{e^{\beta (\omega+i0)}-1}\Big(D_+(\omega) \, e^{2i \omega \lambda} + D_-(\omega) \,e^{-2i \omega \lambda}\Big).\end{align*}
%Where $D_\pm$ are some functions of $\omega$, the explicit form of the relevant coefficient is shown in the equations below. To calculate the integrals of the first and the second term we close the contour in the upper and the lower half-planes, correspondingly. The leading contribution to the integral comes from the pole at $\omega=-i0$:
\begin{align*}
W_\beta(\lambda \to \infty)\approx -\frac{1}{2} \int_{-\infty}^{+\infty} d\omega \frac{1}{\big(e^{\beta (\omega+i0)}-1\big)\sinh \pi (\omega+i0)} e^{-2i\omega \lambda},
\end{align*}
Due to presence of the double pole at $\omega=-i0$ the answer is as follows:
\begin{align*}
W_\beta(\lambda \to \infty)\approx \frac{2\pi}{\beta} \frac{\lambda}{\pi} \approx \frac{2\pi}{\beta} \, W_{BD}(\lambda \to \infty),
\end{align*}
%\textcolor{red}{where $\lambda \sim |\log [t^2 - (x_1 - x_2)^2]| \to %\infty $}.
Note that taking in Eq. \eqref{wightman00} the horizon limit also gives  $W_{\frac{2\pi}{N}}(\lambda \to \infty) \approx N \, W_{BD}(\lambda \to \infty).$

A remarkable fact is the following:  for light--like separations inside the static patch the dominant contribution to the propagator comes from large $\omega$'s; on the contrary, at the horizon small $\omega$'s provide the leading contribution. %That is despite the fact that the horizon is the light--like.
This is because the horizon is the boundary of the patch; the main contribution comes from the infrared rather than ultraviolet frequencies.

%We see that the Wightman propagator with $\beta \neq 2\pi$ has the same structure of the singularity as the Bunch--Davies propagator at the horizon, but with wrong coefficient $\frac{2\pi}{\beta}$. At the same time, the de Sitter invariant Bunch--Davies propagator on the horizon of the static patch has the same singularity as for the light--like separation (\ref{pec}) inside the patch.  Compare the last expression with the one in \eqref{edgebeh}.

\subsection{Flat space limit}

Here we consider the flat space limit, i.e. we let the de Sitter radius go to infinity  $(R\rightarrow\infty)$. Let us start by discussing the flat limit of the modes \eqref{pwholo} and of the BD two-point function, following the treatment given in \cite{bm}. To this aim it is better to use another orbital basis of the forward  lightcone $C^+$:
\begin{equation}
 \xi_+(k) = \left\{\begin{array}{lll}  \xi^0 &=&  \sqrt{k^2 + m^2} /m \cr
\xi^1&=&   k /m  \cr
\xi^2&=& - 1
\end{array}
\right.
\ \ \ \
 \xi_-(k) = \left\{\begin{array}{lll}  \xi^0 &=& \sqrt{k^2 + m^2} /m  \cr
\xi^1&=&    -  k /m   \cr
\xi^2&=& + 1
\end{array}
\right. .
\label{coor2}
\end{equation}
\begin{eqnarray}
\lim_{R\to\infty} \left(\frac{\xi_+(k)\cdot  X\left(\frac {t-i\epsilon} R,\frac x R\right)}{R}\right)^{-\frac{1}{2} - i m R} &=& e^{-i t \sqrt{k^2+m^2} +i k x} \\
\lim_{R\to\infty} \left(\frac{\xi_-(k)\cdot  X\left(\frac {t-i\epsilon} R,\frac x R\right)}{R}\right)^{-\frac{1}{2} - i m R} &=&0
\end{eqnarray}
and so on (recall that  $\mu =  \sqrt{m^2R^2-\frac{1}{4}}$).

%Similarly:
%\begin{multline}
%\lim_{R \rightarrow \infty} \left(\frac{-m\cosh(w)t + mx - mR\sinh(w)}{mR}\right)^{-\frac{1}{2}+imR}
%= \\ =
%\lim_{R \rightarrow \infty}
%\left(-\sinh(w)\right)^{-\frac{1}{2}+imR}
%\left(1+\frac{ m\frac{\cosh(w)}{\sinh(w)} t - m\frac{1}{\sinh(w)} x}{mR}\right)^{-\frac{1}{2}+imR}
%= \\ =
%\lim_{R \rightarrow \infty}  \left(-\sinh(w)\right)^{-\frac{1}{2}+imR}e^{i m\frac{\cosh(w)}{\sinh(w)} t - im\frac{1}{\sinh(w)}x} = \\ =
%\begin{cases}
 %   0 ;\ w<0 \\
  %    \left(-\sinh(w)\right)^{-\frac{1}{2}+\textcolor{red}{imR}}e^{ -i k_0 t - ikx} ;\ \ \ \ w>0
 %\end{cases}
%\end{multline}

It follows that when $R\to \infty$ \cite{bm}
\begin{equation}
 W_{BD}\left(X\left(\frac {t_1-i\epsilon} R,\frac {x_1} R\right), X\left(\frac {t_2+i\epsilon} R,\frac {x_2} R\right)\right)\to \frac 1 {4\pi} \int _{-\infty}^{\infty} e^{-i\sqrt{k^2+m^2} (t_1-t_2-i\epsilon)+ik(x_1-x_2)}  \frac{dk}{\sqrt{k^2+m^2}}\label{GRlim}
\end{equation}
which is the standard Fourier representation of the (positive energy)  Wightman function in Minkowski space.

To  find the flat limit of the Wightman function  $W_\beta$ for  arbitrary $\beta$ in the same we may start by rewriting  the integral representation \eqref{wightman}  in the following way:
\begin{equation}
W_{\beta}(t,x_1,x_2)=\int_{0}^{\infty}d\omega     \left[  e^{-i\omega t} \widetilde  P^R_{\nu}(\omega,x_1,x_2) \frac{1-e^{-2\pi R \omega }}{1-e^{-\beta \omega }}   +    e^{i\omega t}\widetilde  P^R_{\nu}(\omega,x_1,x_2) \frac{1-e^{-2\pi R \omega }}{e^{\beta \omega }-1}\right];
\end{equation}
the superscript $R$ %in $\widetilde P^R_{\nu}(\omega,x_1,x_2)$
indicates  explicitly the restored depende4nce of  (\ref{wightman}) on thde radius $R$.
For $\beta=2\pi R$  the limit $R\to\infty$ (formally) gives
\begin{align}
 \lim_{R\rightarrow\infty} W_{BD}(t,x_1,x_2) = \int_{0}^{\infty}d\omega    e^{-i\omega t} \widetilde P^\infty_{\nu}(\omega,x_1,x_2) .
\end{align}
Taking into account  Eq.  (\ref{GRlim})
it follows that % for arbitrary $\beta$ the limitingexpression for the Wightman propagator can be written as:
\begin{multline}\label{WRinfty}
 \lim_{R\rightarrow\infty}  W_{\beta}(t,x_1,x_2)=\int_{0}^{\infty}d\omega   \left[  e^{-i\omega t} P^\infty_{\nu}(\omega,x_1,x_2) \frac{1}{1-e^{-\beta \omega}}   +    e^{i\omega t}  P^\infty_{\nu}(\omega,x_1,x_2) \frac{1}{e^{\beta \omega }-1}\right]
 = \\=
 \int_{-\infty}^{\infty} \frac{dk}{4\pi \sqrt{k^2 + m^2}}  \left[  \frac{ e^{-i \sqrt{k^2+ m^2} t +i k(x_1-x_2)}}{1-e^{-\beta \sqrt{k^2+m^2} }}   +      \frac{e^{i \sqrt{k^2 + m^2} t-ik(x_1-x_2)}}{e^{\beta \sqrt{k^2 + m^2} }-1}\right]
\end{multline}
%\textcolor{red}{to derive the last equation one should use the relation:\begin{multline*}\widetilde P^\infty_{\nu}(\omega,x_1,x_2) = \int_{-\infty}^{\infty}\frac{dt}{2\pi}\frac {e^{i\omega t}}{4\pi} \int _{-\infty}^{\infty} e^{-i\sqrt{k^2+m^2} (t-i\epsilon)+ik(x_1-x_2)}  \frac{dk}{\sqrt{k^2+m^2}} = \\ = \frac{1}{4\pi}\int_{-\infty}^\infty dk  \frac{e^{ik(x_1-x_2)}}{\sqrt{k^2+m^2}}  \delta(\omega-\sqrt{k^2+m^2}).
%\end{multline*}which follows from\begin{align*}   \widetilde P^\infty_{\nu}(\omega,x_1,x_2) = \lim_{R\to\infty} \int_{-\infty}^{\infty}\frac{dt}{2\pi}e^{i\omega t}W_{BD}(t,x_1,x_2),
%\end{align*}and the large $R$ limit of %$W_{BD}$from \eqref{GRlim}.}
which is precisely the flat space thermal propagator with temperature $1/\beta$. In the  the Bunch-Davies the temperature scales together with $R$ and this maintains invariance at every stage, while in generic case $\beta$ does not scale with $R$. On the other hand  scaling  $\beta = \beta' R$ with constant  $\beta'$ provides in the in the vacuum positive energy Wightman function. This is true also for $\beta = \infty$.

\section{Conclusions and outlook}

Cauchy surfaces in the Rindler -- de Sitter wedge are not Cauchy's for the geodesically complete global de Sitter universe. Giving initial data on such  surfaces completely determines the classical dynamics of fields in the Rindler  -- de Sitter wedge only. By applying the formalism of canonical quantization and Bogoliubov transformations we may construct all the pure Fock states representing quantum Klein--Gordon fields in the wedge. Generalized Bogoliubov transformations \cite{schaeffer,sch2}, however, allow for the construction of a much wider set of states which are, generally speaking, mixed. In this paper we have explicitly constructed all the above states by separating the variables in the static chart (\ref{coordinates}); the construction was exhibited for the two--dimensional de Sitter space not to burden the presentation with unnecessary  complications.

In particular, we gave integral representations of all the KMS states including the Bunch--Davies state at temperature $T=1/2\pi R$. All of them are directly seen to be mixed states, the only pure state in that family being obtained in the zero temperature limit. We also provided explicit formulae for the alpha--states which include also non diagonal terms.

The thermal propagators have unusual pathological singularities on the horizons (vaguely remembering  Einstein's suspicions about the presence of matter on the horizons \cite{janssen}). We mention also that, while these propagators obey the fluctuation--dissipation theorem, the de Sitter invariant Bunch--Davies state, restricted to the wedge, does not possess at least one of the properties of Minkowskian thermal states \cite{Popov:2017xut} because  de Sitter invariance forbids the Debye screening. So there is room for further study.

The important question for cosmology is: what about the initial state of our Universe? The difference between the static patch, the  Poincar\'e patch and the global de Sitter universe  \cite{Akhmedov:2013vka}, \cite{Akhmedov:2019cfd} will  appear in the infrared loops which are sensitive to the initial (and to the boundary conditions).

In flat space--time (at least in a box) an initial  arbitrary state (within a reasonable class)  will thermalize sooner or latter. The temperature of the final state depends on the initial conditions and may be arbitrary. What about thermalization in de Sitter space? Is there thermalization to a state with an arbitrary temperature? How does the answer to these questions depends on the choice of patch (type of initial Cauchy surface)?

To answer these questions one has to resum secularly growing loop corrections. It is the Boltzmann's equation which allows to do that in Minkowski space \cite{Akhmedov:2013vka}. What is the analog of the flat space Boltzmann's equation in the static de Sitter space?

We will address some of the above questions in a forthcoming companion paper.

\section{Acknowledgements}

We would like to acknowledge discussions with A. Semenov, V. Rubakov, A.Polyakov and especially with F. Popov.

The work of ETA was supported by the grant from the Foundation for the Advancement of Theoretical Physics and Mathematics ``BASIS'' and by RFBR grant 18-01-00460. The work of ETA, KVB and DVD is supported by Russian Ministry of education and science (project 5-100).
UM thanks the IHES - Bures-sur-Yvette   where this work has been done and for their generous support during the Covid19 crisis.

\appendix
\section{Appendix}
\label{completeness}
\subsection{Completeness relation of Associated Legendre Functions on the cut}

Here we provide  an explicit (formal) calculation  of the Canonical Commutation Relations  \eqref{cancom}
which, by introducing $\cos\theta = \tanh x = u$,  we rewrite  as follows:

\begin{multline}
 \label{compl2}
 \sin\theta_1  \, \sin\theta_2 \, \delta(\cos\theta_1-\cos\theta_2) =\int_{-\infty}^{\infty} \frac{  \omega \ d\omega }{4\pi \sinh(\pi\mu)}\Gamma\Big(\frac{1}{2}+i\mu-i\omega \Big) \Gamma\Big(\frac{1}{2}- i\mu+ i\omega \Big)  \times \\ \times \Bigg[ \mathsf{P}^{i\omega}_{-\frac{1}{2}+i\mu}(\cos\theta_1)\left(\mathsf{P}^{i\omega}_{-\frac{1}{2}+i\mu}(\cos\theta_2)\right)^*+\mathsf{P}^{i\omega}_{-\frac{1}{2}+i\mu}(-\cos\theta_1)\left(\mathsf{P}^{i\omega}_{-\frac{1}{2}+i\mu}(-\cos\theta_2)\right)^*\Bigg].
\end{multline}
Using the holomorphic plane waves introduced in Sec. (\ref{4444}) we get the following integral representation for $\mathsf{P}^{i\omega}_{-\frac{1}{2}+i\mu}(\cos \theta) $ (see Eq. (\ref{20}) and the following ones):
\begin{equation}
\label{intrep}
\mathsf{P}^{i\omega}_{-\frac{1}{2}+i\mu}(\cos\theta)= \frac{i\, \Gamma(\frac{1}{2}+i\mu)}{2\pi\Gamma(\frac{1}{2}+i\mu-i\omega)}  \int_{-\infty}^{\infty}dt e^{-i\omega t} \Delta f (t,\theta)
\end{equation}
where we set
\begin{eqnarray}
f_{\pm}(t,\theta) &=&  \left(\xi_l(0) \cdot Z(t \pm i\epsilon,\theta )\right)^{-\frac 12 - i \nu}=  \left[\cos\theta+\sin\theta\sinh (t\pm i\epsilon) \right]^{-\frac{1}{2}-i\mu}, \\
\Delta f (t,\theta) &=& \left(  f_+(t,\theta)  - f_-(t,\theta) \right).
\end{eqnarray}
$\mathsf{P}^{i\omega}_{-\frac{1}{2}+i\mu}(\cos \theta) $ is therefore the Fourier transform of the discontinuity of the holomorphic plane waves on the real de Sitter manifold.
Let us insert  \eqref{intrep} in Eq. \eqref{compl2}; let us  consider for instance the first term on the rhs of Eq. \eqref{compl2}. By performing the integration over $\omega$ we get
%\begin{align*}\int_{-\infty}^{\infty} dt_1 dt_2d\omega \omega e^{i\omega(t_2-t_1)}g(t_1)f(t_2)=-\pi i\int dt \Big( g'(t)f(t)-g(t)f'(t) \Big),\end{align*}
\begin{eqnarray}
&& {(\rm \ref{compl2})} =  -\frac{i }{16\pi \sinh^2 \pi\nu } \int_{-\infty}^{\infty} dt\left[
 \left(\partial_{t} \Delta f (t,\theta_1)  \right)
\Delta f (t,\theta_1)^*-
  \Delta f (t,\theta_1)
   \partial_{t} \Delta f (t,\theta_2)^* \right]+  \cr
  &&- \frac{i }{16\pi \sinh^2 \pi\nu } \int_{-\infty}^{\infty} dt\left[
 \left(\partial_{t} \Delta f (t,\pi -\theta_1)  \right)
\Delta f (t,\pi - \theta_1)^*-
  \Delta f (t,\pi - \theta_1)
   \partial_{t} \Delta f (t,\pi - \theta_2)^* \right] = \cr &&\cr  &&\cr
  && =   -\frac{i }{16\pi \sinh^2 \pi\nu }\sum_{k=-\pm} \int_{-\infty}^{\infty} dt\left[
 \left(\partial_{t}  f_k (t,\theta_1)  \right)
 f_ k(t, \theta_1)^*-
   f_k (t,\theta_1)
    \partial_{t}  f_k (t, \theta_2)^* \right] +
    \cr
   && -\frac{i }{16\pi \sinh^2 \pi\nu }\sum_{k=-\pm} \int_{-\infty}^{\infty} dt\left[
 \left(\partial_{t}  f_k (t,\pi- \theta_1)  \right)
 f_ k(t,\pi - \theta_1)^*-
   f_k (t,\pi - \theta_1)
    \partial_{t}  f_k (t,\theta_2)^* \right]. %+
\end{eqnarray}
In the second step we used the analyticity properties of the plane waves; this simplification is valid  in the two-dimensional spacetime and in any even dimensional spacetime as well.
By introducing  the Mellin representation of the plane wave:
\begin{align}
\label{intrep2}
f_{\pm}(t,\theta) =\frac{e^{\mp \frac{i\pi}{2}   (\frac 12 + i \nu)}}{\Gamma( \frac 12 + i \nu)}\int_0^\infty du \ u^{-\frac{1}{2}+i\nu}
e^{\pm iu(\cos \theta + \sin\theta \,  \sinh (t\pm i\epsilon) )},\ \ \ \ \  0<\theta<\pi,
\end{align}
a few easy integrations show the validity of Eq. (\ref{compl2}) and the completeness of the modes.

\end{document}